\documentclass{article}
\usepackage{graphicx} 
\usepackage{amsmath}

\usepackage[a4paper, total={6in, 8in}]{geometry}

\usepackage[style=vancouver]{biblatex} 
\usepackage{xurl} 
\usepackage[hidelinks]{hyperref}
\usepackage{placeins} 
\let\Oldsection\section
\renewcommand{\section}{\FloatBarrier\Oldsection}
\let\Oldsubsection\subsection
\renewcommand{\subsection}{\FloatBarrier\Oldsubsection}
\let\Oldsubsubsection\subsubsection
\renewcommand{\subsubsection}{\FloatBarrier\Oldsubsubsection}

\usepackage{makecell}
\setcellgapes{5pt}
\makegapedcells

\usepackage{authblk}
\title{A method for including additional socio-demographic factors in social contact matrices for compartment-based epidemic models}
\author[1,2]{Vincent X. Lomas}
\author[2]{Tim Chambers}
\author[1]{Leighton M. Watson}
\author[1,3]{Michael Plank}
\affil[1]{School of Mathematics and Statistics, University of Canterbury, Christchurch, Aotearoa}
\affil[2]{Ng\=ai Tahu Research Center, University of Canterbury, Christchurch, Aotearoa}
\affil[3]{Te P\=unaha Matatini Centre of Research Excellence in Complex Systems, Auckland, Aotearoa}

\begin{document}

\maketitle

\section{Abstract}
Socio-demographic factors influence social contact patterns and play a fundamental role in shaping the transmission dynamics of infectious diseases. Compartment-based models of infectious disease dynamics commonly consider the dependence of contact patterns on age, but ignore other factors that are likely to have compounding effects. Methods that stratify the population by multiple socio-demographic factors are few and require social contact surveys that contain information about all factors of interest. Here we present a method that can stratify an existing social contact matrix with an additional socio-demographic factor using information about the population structure of the socio-demographic factors and assumptions about the aggregate mixing rates within and between groups. We then analyse hypothetical populations and a projection of a social contact survey onto Aotearoa New Zealand's age-ethnicity structure to show how these extended social contact matrices can change epidemic dynamics and outcomes. The inclusion of the additional factor has a big impact on the model reproduction number and final epidemic size. We find that minority group epidemic outcomes are most sensitive to variation in model parameter values.

\section{Introduction}
Mathematical models of infectious disease spread help to inform policy decisions about public health interventions. Contact matrices are important inputs into these infectious disease models that allow group interaction rates to be aggregated from the frequency, length, and likelihood of infection from individual interactions. These group interaction rates allow us to account for the fact that socially active individuals are more likely to be infected and, when infected, will spread the disease to more people \cite{britton_mathematical_2020}. They also allow us to better analyse how different socio-demographic groups are affected by disease spread.

There are many choices for which socio-demographic groups to consider, but a common choice is age groups \cite{naidoo_incorporating_2024}. The nature and extent of social interactions can differ substantially based on age, with younger people generally having larger social networks and higher risk interactions, than older people \cite{marcum_age_2013}. Models that include these age-based social contact matrices have been shown to better capture infectious disease dynamics \cite{wallinga_using_2006}.

Despite the known importance of age, it is not the sole socio-demographic factor that drives disease spread. In some circumstances, within age group heterogeneity can be more important than between age group heterogeneity \cite{britton_improving_2025, kuchel_workshop_2025}. Models that only consider age can fail to capture social and behavioural heterogeneities; lead to preventable socio-economic, racial, and geographic health disparities; and can obscure important variation in disease outcome \cite{bedson_review_2021,zelner_there_2022,tizzoni_addressing_2022}. There has been less consideration of socio-demographic status in social contact matrices and its potential impact on health outcomes \cite{bambra_pandemic_2022}. Even when socio-demographics characteristics other than age are included in models, they are rarely a central consideration and are often under-utilised \cite{tizzoni_addressing_2022}. As a result, these models can often bias results towards average interaction rates and do not fully capture the full risk of some socio-demographic groups. If modelling results are biased and used to inform policy decisions, this can introduce or further exacerbate existing inequities in health outcomes \cite{debruin_social_2012,zipfel_health_2021}. There is a clear need for models that explicitly consider a greater range of socio-demographic factors to better account for heterogeneities in social and behavioural dynamics.

One simple way to integrate socio-demographic factors into models is to consider how each factor by itself affects disease spread. Considering each factor alone can fail to observe synergistic effects such as deprivation affecting comorbidity or different levels of deprivation having different age structures \cite{goodfellow_covid-19_2024,zipfel_health_2021}, and might lead to underestimation of the basic reproductive number and the impact on certain groups of an infectious disease \cite{manna_importance_2024,manna_generalized_2024}. We need to not only consider a wider scope of socio-demographic heterogeneities, but also study how the intersection of these socio-demographic factors affects disease spread. To be able to consider multiple socio-demographic factors, estimates of the social contact rates between each combination of socio-demographic factors are needed.

Methods to extend social contact matrices to consider additional socio-demographic factors are limited; the most relevant of which explored generalized contact matrices using an assortative mixing method and their effect on herd immunity and adherence to non-pharmaceutical interventions \cite{manna_generalized_2024}. This work allowed for the construction of an extended contact matrix with a given age structure and group-specific assortativity values. However, this method required empirical information (like estimates from social contact surveys or similar) on social contact rates and did not allow for specification of the effect the additional group had on otherwise identical individuals, which will be discussed further in our methods. Another method considered three socio-demographic effects (and the effect of neighbourhood levels) \cite{domenico_individual-based_2025}. This work estimated contact matrices stratified by individual-based and area-based socio-demographic factors using a social contact survey (the European CoMix survey) in Sweden. They did this in a way that did not require socio-demographic information about the individuals the survey participants contacted, however they did require that information from the survey participants. Information derived from empirical studies is time-consuming and expensive to collect. Thus, models reliant on these data have reduced generalisability. Directly estimating pairwise interaction rates using social contact surveys can also be difficult as perception bias present in data collection can lead to large inaccuracies \cite{harris_simulating_2025}.

Here, we create a method that can stratify an existing social contact matrix with an additional socio-demographic factor. We present a method that gives a principled way to construct extended matrices that allows for group-based assortativity and group-dependent contact rates and in situations where detailed information on how socio-demographic  groups interact with one another is lacking. Instead we use information on the population structure of socio-demographic groups; the relative contact rate, and a measure of the new socio-demographic groups' assortativity (or how much groups preferentially interact within themselves). Firstly we outline a set of three conditions we require any social contact matrix to satisfy. Secondly, we present a solution to these conditions for proportionate and assortative mixing. Finally, we explore numerical simulations of an SEIR (Susceptible, Exposed, Infectious, and Recovered) models with this contact matrix extension to highlight obscured variation in epidemic outcomes that our method reveals to highlight potential variation in disease outcomes estimated by our method relative to single-factor models.


\section{Methods}

Suppose we have a population stratified by some socio-demographic factor with $n_A$ groups with a known contact matrix, $C_{ij}$, that represents the average number of contacts each day an individual in socio-demographic group $i$ has with individuals in socio-demographic group $j$ ($i, j = 1, . . . , n_A$). In practice, the socio-demographic factor will most commonly be age and $C_{ij}$ will be an age-based contact matrix, but this is not a requirement and our method is applicable to other socio-demographic factors. Then the equations that model the spread of the disease for each group of the population are given by the following:

\begin{align*}
    \frac{dS_j}{dt} & = -qs_j\frac{S_j}{N_j}\sum_i C_{ij}I_i\\
    \frac{dE_j}{dt} & = qs_j\frac{S_j}{N_j}\sum_i C_{ij}I_i - \sigma E_j\\
    \frac{dI_j}{dt} & = \sigma E_j - \gamma I_j\\
    \frac{dR_j}{dt} & = \gamma I_j,
\end{align*}
where $\sigma$ is the rate of disease development; $\gamma$ is the rate of disease recovery; $S_j$, $E_j$, $I_j$, and $R_j$ are vectors containing the susceptible, exposed, infectious, and recovered individuals in group $j$, respectively; $N_j$ is the population in group $j$; $q$ is the probability of transmission given a contact between an infectious and a susceptible individual; and $s_j$ is the average susceptibility of individuals in socio-demographic group $j$.

The social contact matrix has a symmetry condition to ensure the two-way nature of contacts; that being that if someone is interacting with another person, that other person must also be interacting with them.

\begin{align}
    N_iC_{ij} = N_jC_{ji}, \label{eq:old_symm}
\end{align}

where $N_i$ is the population size in socio-demographic group $i$. The next generation matrix (in a fully susceptible population) can be found in terms of this social contact matrix and is given by the following:

\begin{align*}
    K_{ij} = q C_{ij}s_j/\gamma,
\end{align*}

where the basic reproduction number $R_0$ is the dominant eigenvalue of the next generation matrix ($K_{ij}$). Typically, the value of $q$ is chosen to give a pre-specified value of $R_0$.

Now suppose we want to additionally stratify the population by a second socio-demographic factor with $n_D \geq 2$ groups, resulting in a total of $n_An_D$ groups. In principle, these could represent any two socio-demographic factor, but form here on, for clarity, we will refer to the original socio-demographic factor as age and the additional factor as ethnicity. To parametrise a model such as this, we need an extended contact matrix $C_{ia,jb}$ representing the average number of contacts an individual in age group $i$ and ethnicity group $a$ has with individuals in age group $j$ and ethnicity group $b$, where $a$ and $b$ refer the ethnicity groups of the individuals ($a,b = 1, . . . , n_D$). Such a matrix would allowing for modelling the spread of disease stratified by the two socio-demographic factors with the following equations:

\begin{align*}
    \frac{dS_{jb}}{dt} & = -qs_{jb}\frac{S_{jb}}{N_{jb}}\sum_{i,a} C_{ia,jb}I_{ia}\\
    \frac{dE_{jb}}{dt} & = qs_{jb}\frac{S_{jb}}{N_{jb}}\sum_{i,a} C_{ia,jb}I_{ia} - \sigma E_{jb}\\
    \frac{dI_{jb}}{dt} & = \sigma E_{jb} - \gamma I_{jb}\\
    \frac{dR_{jb}}{dt} & = \gamma I_{jb},
\end{align*}
where $S_{jb}$ is the susceptible population of people belonging to age group $j$ and ethnicity group $b$ (similarly for $E$, $I$, and $I$); $N_{jb}$ is the population of people belonging to age group $j$ and ethnicity group $b$; and the average susceptibility of individuals belonging to age group $i$ and ethnicity group $b$ is $s_{jb}$. Here, and in future equations, we will use the notation $\sum_{i,a}$ to mean summation over all $i$ and all $a$, or equivalently $\sum_{i=1}^{n_A}\sum_{a=1}^{n_D}$, where the bounds of summation match the range of possible values being summed over.

\subsection{Required properties of the extended contact matrix}

The new social contact matrix must satisfy the constraint that when aggregating over ethnicity, contacts between individuals in age groups $i$ and $j$ sum to the original contact matrix element $C_{ij}$. This condition ensures that extensions of the contact matrix do not add nor remove contacts and instead distribute them by ethnicity.

\begin{align}
    \frac{\sum_{a,b}N_{ia}C_{ia,jb}}{N_i} = C_{ij}, \label{eq:aggregation_cond}
\end{align}
where $N_{ia}$ is the population of people belonging to both age group $i$ and ethnicity group $a$ and $N_i=\sum_aN_{ia}$ is the population of people belonging to age group $i$. For convenience, we will overload this notation by using $N_a=\sum_iN_{ia}$ to denote the population of ethnicity group $a$.

The previously outlined symmetry condition (Equation \ref{eq:old_symm}) must also hold for this new matrix to conserve the two way nature of contacts.

\begin{align}
    N_{ia}C_{ia,jb} = N_{jb}C_{jb,ia}. \label{eq:symmetry_cond}
\end{align}

Additionally, we would like to be able to specify ethnicity-specific interaction frequencies $F_a$ (which we will refer to as relative contact rates). In the absence of direct information on pairwise contact rates $C_{jb,ia}$, we assume that these relative contact rates apply equally across all age groups. In other words, we require the ratio of the total contact rate in age group $i$ and ethnicity group $a_1$ to the total contact rate in age group $i$ and ethnicity group $a_2$ to be $F_{a_1}/F_{a_2}$, for all $i=1,\ldots,n_A$. We will henceforth refer to this condition as the ``relative interaction condition". 

\begin{align}
    \frac{\sum_{j,b}C_{ia_1,jb}}{\sum_{j,b}C_{ia_2,jb}} = \frac{F_{a_1}}{F_{a_2}}, \qquad i=1,\ldots,n_A. \label{eq:contact_ratio_cond}
\end{align}

This condition provides a novel way to extend existing contact matrices to include an additional factor when there is group-specific information about total contact frequencies, but not about all group pairwise contact rates.

\subsection{Proportionate mixing between ethnicity groups}
We first explore the simplest case that assumes each age group's contacts are distributed across the ethnicity groups in proportion to their population size and relative contact rates. This can be modelled via the following definition for $C_{ia,jb}$:

\begin{align} \label{eq:prop_matrix}
    C_{ia,jb}^{\text{proportionate}}& = \frac{F_aF_bN_{jb}C_{ij}}{\bar F_i\bar F_jN_j},
\end{align}
where we define $\bar F_i={\sum_a F_aN_{ia}}/{N_i}$ to represent the average contact rate in age group $i$. Proof of satisfaction of conditions for this matrix is present in Appendix A. We can gain some intuition behind this matrix by observing that it distributes the total number of contacts between age groups $i$ and $j$, $C_{ij}N_i$, across ethnicity groups in proportion to the fraction of age groups $i$ and $j$ in those groups ($N_{ia}/N_i$ and $N_{jb}/N_j$ respectively) and their relative contact rates ($F_a$ and $F_b$) (see Appendix A for further detail).

\subsection{Assortative mixing between ethnicity groups}
We now consider assortativity, which is the tendency of people to preferentially mix with people in their own ethnicity group. We first consider the limiting case where ethnicity groups interact only within their own ethnicity group, i.e. all contacts occur within a ethnicity group, which we will call segregated mixing.

Our three conditions do not uniquely define a segregated-mixing matrix and there are infinitely matrices that satisfy them. This is a consequence of the fact that the number of non-zero matrix elements is greater than the number of independent constraints. We present a family of solutions for segregated mixing. Let $G_{ija}$ be a function that satisfies the following two conditions: $G_{ija}=G_{jia}$ and $\sum_a G_{ija} = 1$, where $i$, $j$, and $a$ denote the respective socio-demographic groups and match $C_{ia,jb}$. The segregated mixing matrix can then be represented as the following:
\begin{align}
    C^{\text{segregated}}_{ia,jb}&= \begin{cases}
        \delta_{ab}\left[\frac{F_a}{\bar F_i} \sum_{k} C_{ik} - \frac{N_i}{N_{ia}} \sum_{k\neq i} C_{ik} G_{ika}\right] &\text{if } i=j,\\
        \delta_{ab}\frac{N_iC_{ij}}{N_{ia}}  G_{ija} & \text{otherwise},
    \end{cases}\label{eq:gen_seg_matrix}
\end{align}
Proof of satisfaction of conditions for this matrix is present in Appendix A. We select the solution that distributes contacts between different age groups $i$ and $j$ across ethnicity groups in proportion to the mean of the fraction of age group $i$'s contacts that are in ethnicity group $a$ and the fraction of age group $j$'s contacts that are in ethnicity group $a$. Within age group mixing is then defined to satisfy the conditions. This results in the following solution for $C_{ia,jb}$:

\begin{align}
    G_{ija}&=\frac12\left(\frac{F_aN_{ia}}{\bar F_iN_i}+\frac{F_aN_{ja}}{\bar F_jN_j}\right)\\
    C^{\text{segregated}}_{ia,jb}&= \begin{cases}
        \delta_{ab}\left[\frac{F_aC_{ij}}{\bar F_i} +\sum_{k\neq i}\frac{F_aC_{ik}N_i}{2N_{ia}}\left(\frac{N_{ia}}{\bar F_iN_i}-\frac{N_{ka}}{\bar F_kN_k}\right)\right] &\text{if } i=j,\\
        \delta_{ab}\frac{F_aC_{ij}N_i}{2N_{ia}}\left(\frac{N_{ia}}{\bar F_iN_i}+\frac{N_{ja}}{\bar F_jN_j}\right) & \text{otherwise},
    \end{cases} \label{eq:seg_matrix}
\end{align}
where $\delta_{ab}$ is 1 if $a=b$ and 0 otherwise. Some other possible choices of $G_{ija}$ are present in Appendix A.

Depending on the choice of $G_{ija}$ and the extremity of parameters/group size differences, the extended contact matrix can become negative for some within-age-group interaction rates. In realistic scenarios we do not expect this to occur, however, when applying this method, verification of positive matrix elements is required.

We now introduce the ethnic assortativity constant, $\epsilon$, which is the degree to which ethnicity groups prefer to interact within their own ethnicity. We define the assortative mixing matrix as a linear combination of the segregated mixing and proportionate mixing contact matrices where $\epsilon$ is the proportion of assortativity in the interaction for the ethnicity groups ($\epsilon=0$ is purely proportionate mixing, while $\epsilon=1$ is purely segregated).

\begin{align*}
    C_{ia,jb}^\text{assortative} = (1-\epsilon)C^{\text{proportionate}}_{ia,jb} +\epsilon C^{\text{segregated}}_{ia,jb}
\end{align*}
Since $C^{\text{proportionate}}$ and $C^{\text{segregated}}$ satisfy the required conditions by construction, and the conditions are all linear, it follows that $C^{\text{assortative}}$ also satisfies the required conditions.

\subsection{Contact matrix without the age group}
We can aggregate our contact matrix over the age groups to recover a contact matrix in terms of interactions between the ethnicity groups:

\begin{align*}
    C_{ab} = \frac{\sum_{i,j}N_{ia}C_{ia,jb}}{N_a},
\end{align*}

which also satisfies the conditions. This can allow us to estimate the ethnicity-structured social contact matrix by using the age social contact matrix and information about how age and ethnicity structures relate to each other.

Note that it is possible to re-index the four-dimensional array $C_{ia,jb}$ into a two-dimensional matrix $\tilde{C}_{i'j'}$ indexed by $i'$ and $j'$ where each combination of $i$ and $a$ is assigned to an integer and $i'=1,2,...,n_An_D$. The re-indexed matrix can be further extended with an addition socio-demographic factor using the same process as for $C_{ij}$. In theory this allows social contact matrices to be extended with an arbitrary amount of socio-demographic factors. However, each time the contact matrix is extended we need to make additional assumptions about the relationships between socio-demographic factors which can limit the number of realistic extensions made. The number of compartments the model requires would also quickly become infeasible.

\subsection{Special cases}

If we let the ethnicity groups have equal relative contact rates (i.e. $F_a=F_b$ for all $a$ and $b$), the contact matrices reduce to the following forms:

\begin{align*}
    C_{ia,jb}^{\text{proportionate}} & = \frac{C_{ij}N_{jb}}{N_j}\\
    C^{\text{segregated}}_{ia,jb} &= \begin{cases}
        \delta_{ab}\left[C_{ij} +\sum_{k\neq i}\frac{C_{ik}N_i}{2N_{ia}}\left(\frac{N_{ia}}{N_i}-\frac{N_{ka}}{N_k}\right)\right] \quad & \text{if } i=j,\\
        \delta_{ab}\frac{C_{ij}N_i}{2N_{ia}}\left(\frac{N_{ia}}{N_i}+\frac{N_{ja}}{N_j}\right) & \text{otherwise}
    \end{cases}
\end{align*}

If instead we let all ethnicity groups have the same age group distribution (but not necessarily the same total population), then $N_{ia}=N_iN_a/N$ and equations the social contact matrix equations reduce to the following forms:

\begin{align*}
    C_{ia,jb}^{\text{proportionate}} 
    &= \frac{F_aF_bC_{ij}N_b}{\bar F^2N}\\
    C^{\text{segregated}}_{ia,jb} &=
        \delta_{ab}\frac{F_aC_{ij}}{\bar F_i},
\end{align*}
where $\bar F = \sum_a F_aN_a/N$ is the average population-wide contact rate.

If we let the ethnicity groups have equal relative contact rates and the same age group distribution, then the contact matrices further reduce to the following forms

\begin{align*}
    C_{ia,jb}^{\text{proportionate}} & = \frac{C_{ij}N_{b}}{N}\\
    C^{\text{segregated}}_{ia,jb} &= \delta_{ab}C_{ij}
\end{align*}

\subsection{Numerical simulation method}
We analysed numerical simulations of the epidemic model when a social contact matrix was extended with ethnicity. We first analysed hypothetical populations and contact matrices, and then analysed an European empirical social contact matrix (the POLYMOD study \cite{mossong_social_2008}, a social contact survey conducted in many European countries) projected onto Aotearoa New Zealand, both methods are outline below. We then ran SEIR simulations using the populations and extended social contact matrix.

\subsubsection{Hypothetical populations}
We constructed various hypothetical population distributions and parameter values in different scenarios (seen in Table \ref{tab:scenario_table}). For all scenarios, we construct a simple 5 age group social contact matrix to be used as the base for the extended matrix via the following formula,

\begin{table}[ht]
    \centering
    \begin{tabular}{c|c|c|c}
        Scenario & Group sizes ($N_a$) & Group age structures ($N_{ia}/N_i$) & Age contact rates ($\sum_jC_{ij}$)\\
        \hline
        1 & Equal & Equal & Equal\\
        \hline
        2 & Equal & Different & Equal\\
        \hline
        3 & Different & Equal & Equal\\
        \hline
        4 & Different & Different & Equal\\
        \hline
        5 & Different & Different & Different
    \end{tabular}
    \caption{Table of parameter values for the 5 scenarios considered. Here ``groups" refers to the ethnicity groups. When ethnicity groups were different sizes, ethnicity group 1 made up 90\% of the total population and ethnicity group 2 made up 10\%. When the age structure was the same, each ethnicity group had equal populations in every age group. When they were different, there was a linear relationship between age and population with the youngest age being half the population of the oldest group in ethnicity group 1 and vice-verse for ethnicity group 2. Full descriptions of the scenarios can be found in the Appendix B.}
    \label{tab:scenario_table}
\end{table}

\begin{align*}
    C_{ij} = (1-c) \frac{a_ia_j}{\sum_k a_kN_k} + \delta_{ij}c \frac{a_i}{N_i}.
\end{align*}

Where $c=0.3$ is the age-assortativity, or percent of contacts within an individual's age group before proportionate mixing was applied (note that after proportionate mixing is applied, the proportion of contacts within-group would be greater) and $a_k$ and $N_k$ are age group $k$'s contact rate and population, respectively. 

\subsubsection{POLYMOD projection}
Using the R package Conmat \cite{tierney_conmat_2026} and age-ethnicity population counts (those ethnicity groups being Māori, Pacific, Asian, and European/Other separated by 5 year ages bands and group everyone above 90 years old) in Aotearoa New Zealand during 2023 from StatsNZ \cite{statsnz_2023_2023}, the POLYMOD study's contact matrix was projected onto Aotearoa New Zealand. We then required that this matrix was symmetric via the following equation: $C_{ij}^{\text{symmetric}}=(C_{ij}+(N_iC_{ij})^T/N_j)/2$. We then extended the resulting matrix using total ethnicity group populations by age with relative contact rates of 2, 3, 0.9, and 1 for the Māori, Pacific, Asian, and European/Other groups, respectively, and an ethnic assortativity value of $0.2$. These parameter estimates are used here for illustrative purposes, and were taken from our previous study of ethnicity-specific transmission dynamics of Covid-19 in New Zealand in 2022 \cite{lomas_modelling_2025}. This previous work found the average difference between the contact rates of the whole population while parameters $F_a$ used here quantify the average difference between the contact rates of ethnicity groups within an age group.

\section{Numerical simulation results}
We present our results for the numerical simulations of the epidemic model using the extended social contact matrices. We first present the hypothetical population example and then the empirical projection example.

\subsection{Hypothetical population results}

\begin{figure}[ht]
    \centering
    \includegraphics[width=0.95\linewidth]{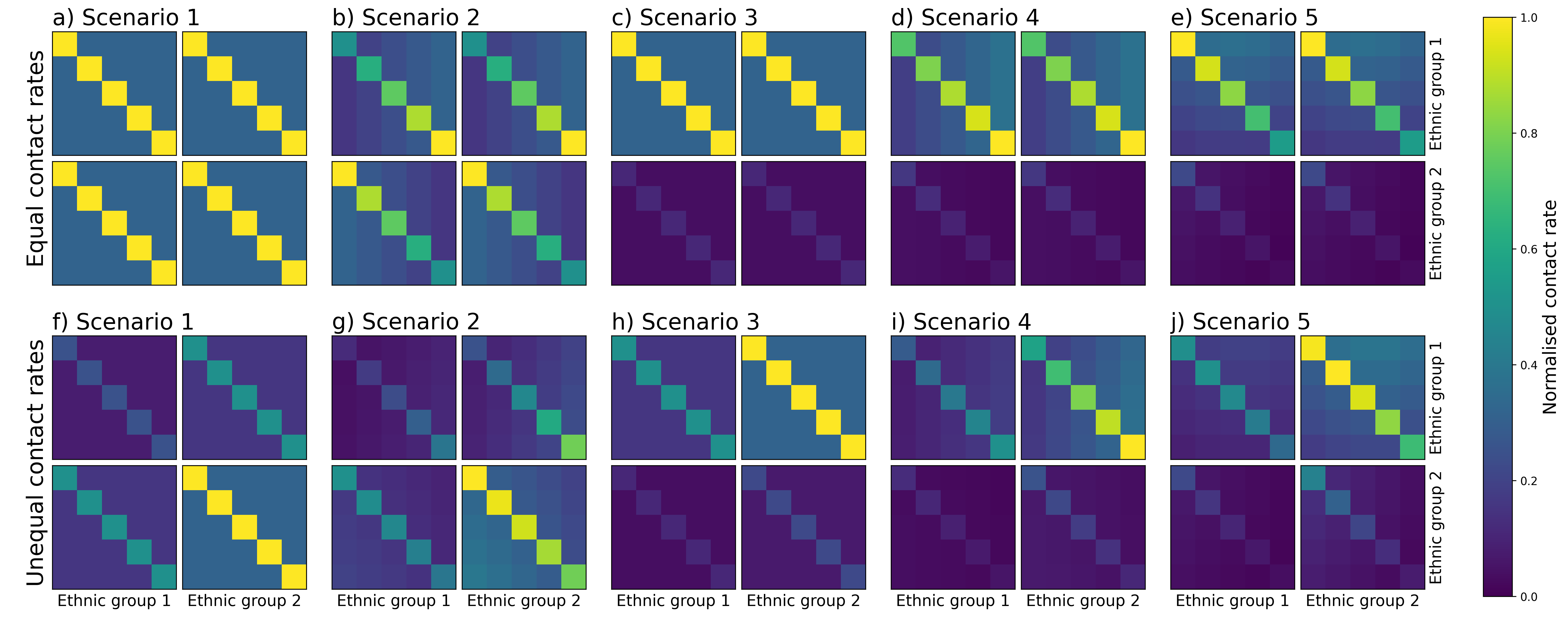}
    \caption{Heat-plots of the purely proportionate mixing contact matrices for each scenario; plots show scenarios 1-5 from left to right with the same relative contact rates in plots a-e, while ethnicity group 1 had the half the relative rate of ethnicity group 2 in plots f-j. Contact matrices are normalised such that the largest value in each is 1. The top-left quadrant is the interaction of age groups within ethnicity group 1, the bottom right is the same for ethnicity group 2, and the off-diagonal quadrants are the interactions between ethnicity groups.}
    \label{fig:heatplot_eth0}
\end{figure}

\begin{figure}[ht]
    \centering
    \includegraphics[width=0.95\linewidth]{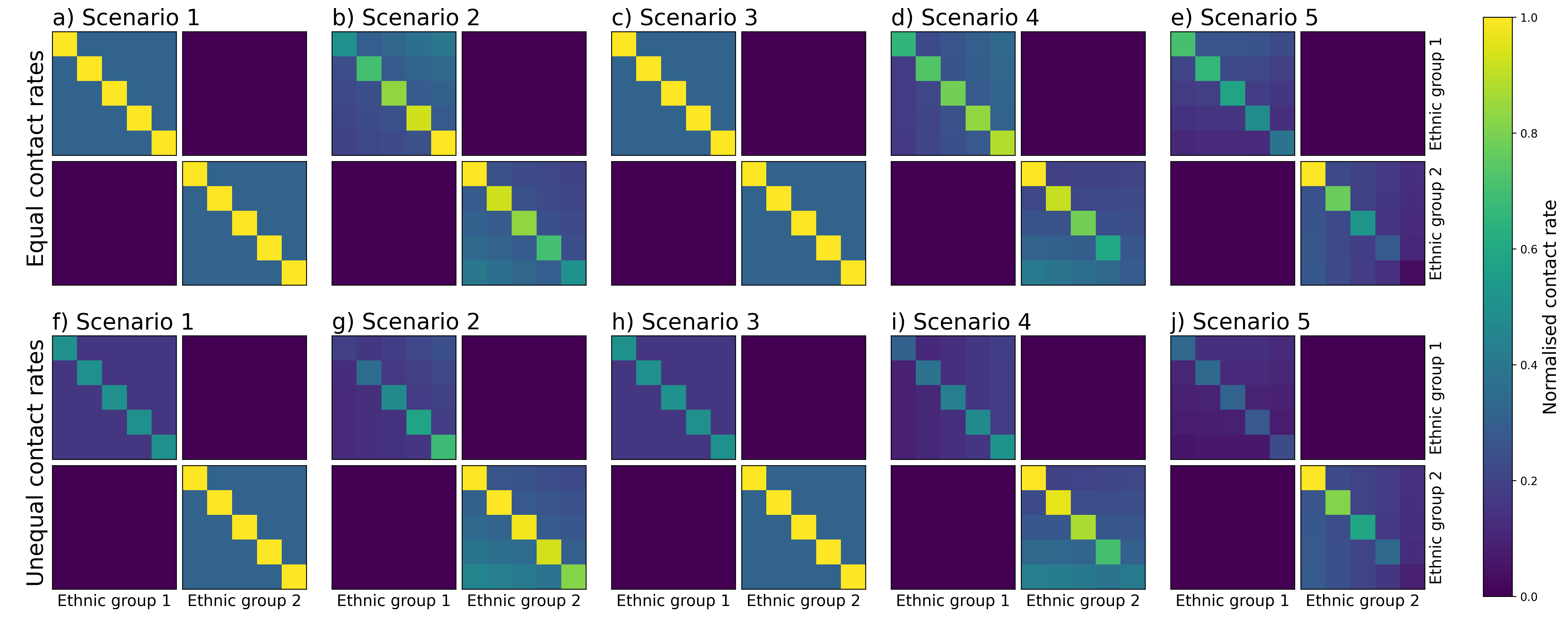}
    \caption{Heat-plot of the segregated mixing contact matrices for each scenario; plots show scenarios 1-5 from left to right with the same relative contact rates in plots a-e, while ethnicity group 1 had the half the relative rate of ethnicity group 2 in plots f-j. Contact matrices are normalised such that the largest value in each is 1. The top-left quadrant is the interaction of age groups within ethnicity group 1, the bottom right is the same for ethnicity group 2, and the off-diagonal elements are the interactions between ethnicity groups.}
    \label{fig:heatplot_eth1}
\end{figure}

Normalised heat-plots of the ethnicity group extension of the contact matrix for all scenarios can be seen in Figures \ref{fig:heatplot_eth0} and \ref{fig:heatplot_eth1}, respectively. These plots show that when the population structure is the same, each within-ethnicity interaction block is a (potentially scaled) copy of the age only matrix. Different population structures lead to horizontal and vertical loading on the (proportionally) larger age groups within each. We numerically verified that these extended contact matrices satisfied the three conditions defined in Equations \ref{eq:aggregation_cond}-\ref{eq:contact_ratio_cond}.

\subsubsection{Relative ethnicity group contact rate variation:}

We show final attack rates (or the proportion of individuals infected during the epidemic) from a simulated epidemic by two aggregation levels: the ethnicity group level to show the differences by ethnicity group unobtainable from purely age-structured models; and the whole population level to show variation in population-wide dynamics which are obscured without the inclusion of ethnicity in the model.

\begin{figure}[ht]
    \centering
    \includegraphics[width=0.95\linewidth]{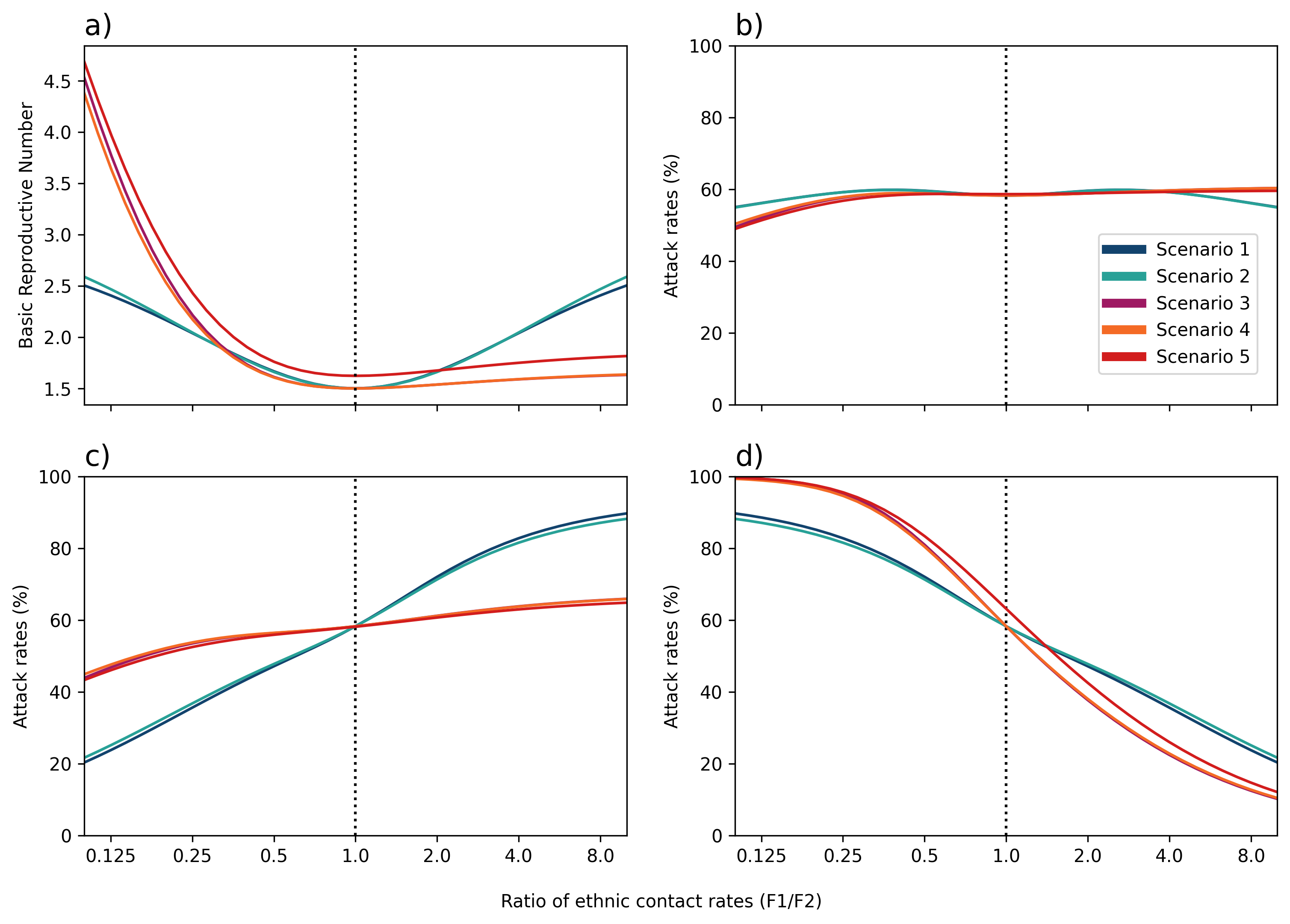}
    \caption{Plots of the variation in a) the basic reproductive number; b) the whole population attack rates; c) ethnicity group 1's attack rates; and d) ethnicity group 2's attack rates with relative ethnicity contact rate. A lower relative ethnicity contact ratio means the ethnicity group with a lower population (if present) has the higher contact rate. The ethnic assortativity is 0.3 for this figure.}
    \label{fig:rel_eth_var}
\end{figure}

Figure \ref{fig:rel_eth_var} shows how $R_0$ and attack rate in a simulated epidemic varies with the contact rate ratio. The dotted line at 1 shows the point corresponding to the model that does not take ethnicity into account. Note that at the dotted line, scenario 5 shows slight differences in attack rate compared to the other scenarios due to different age-specific contact rates (Table \ref{tab:scenario_table}). Here we show variation in the contact rate ratio between $0.1$ and $10$. A realistic value for this contact rate ratio would likely fall within this range as previous work \cite{ma_modeling_2021,lomas_modelling_2025} has estimated population-wide contact rate ratios in the range $0.2-5$. However, more extreme differences may be possible in some contexts. In part a) of the figure, we see the basic reproductive number increasing in all scenarios as we deviate from a contact rate ratio of 1. As the contact rate of one group increases, it results in that group being both more likely to get infected and, when infected, spread the disease more. This increase in the basic reproduction number is greater than the decrease caused by the reduction of the other group's contact rate, which leads to an overall increase in the basic reproductive number. The rate of increase is determined by the population structure of the ethnicity groups. Scenarios 1-2 see a symmetric difference due to the similar population sizes between groups, while scenarios 3-5 see a large increase as the smaller group has more contacts and a small increase in the opposite direction. The magnitude of increase depends on the proportion of each age group the ethnicity group makes up, i.e. if an ethnicity group makes up half the population of an age group, that group's contact rates can at most double and the reproductive number (if that was the only age group) would double as well. The individual ethnicity group attack rates increase with an increase in their contact rate in a way that is initially roughly proportional to the inverse of their population, but tapers off near 0\% and 100\%. In scenarios 1 and 2, the rate of attack rate increase for the group with an increasing contact rate is greater than the rate of decrease for the other group leading to the overall attack rate increasing in a region near a contact rate ratio of 1. When the group with an increasing contact rate approaches saturation, the other group will see a faster decrease. This can be seen in scenarios 3-5 (near a relative contact ratio of 0.2) on Figure \ref{fig:rel_eth_var}.

Note that the fact that $R_0$ is an increasing function of the difference between group-specific contact rates is an expected outcome, consistent with well-known results that increasing contact heterogeneity in a model increases $R_0$ \cite{diekmann_definition_1990,may_infection_2001}. This happens because $R_0$ is the expected number of secondary infections caused by an average infected individual, and in models with heterogeneity, infected individuals are disproportionately likely to be those with high contact rates. In practice, the model may be calibrated by choosing the transmission coefficient $q$ to give a desired value of $R_0$ or to fit with empirical observations. To investigate the consequences of heterogeneity in this situation, Appendix C shows results when the extended social contact matrix is scaled such that $R_0$ remains constant as relative contact rate varies. 

\subsubsection{Transmission probability variation}

\begin{figure}[ht]
    \centering
    \includegraphics[width=0.95\linewidth]{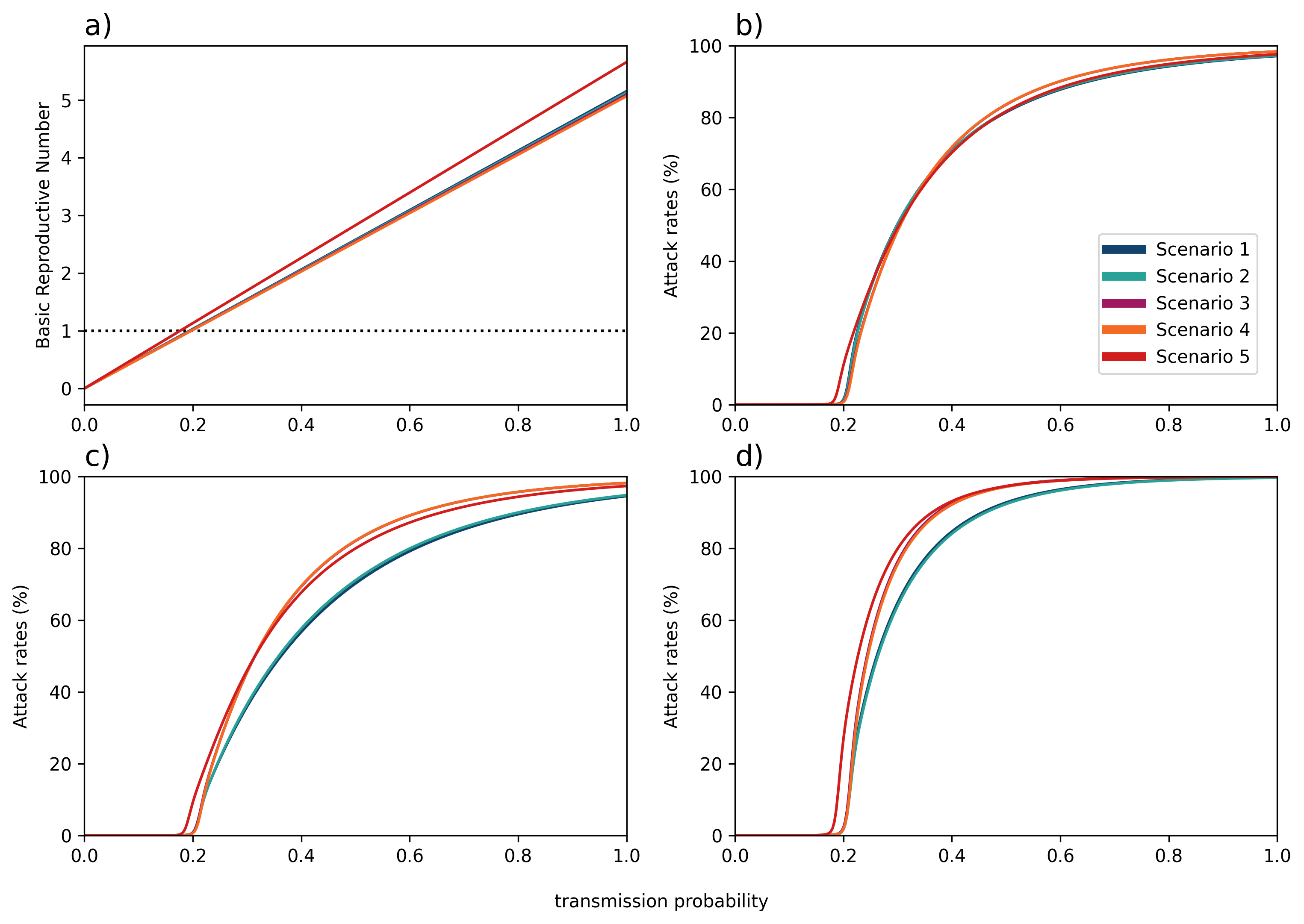}
    \caption{Plots of the variation in a) the basic reproductive number with a dotted line at $R_0=1$; b) the total population attack rates; c) ethnicity group 1's attack rates; and d) ethnicity group 2's attack rates with transmission probability. ethnic assortativity is 0.3 and relative contact rates are $[1,2]$, i.e. the ethnicity group with a lower population (in scenarios 3-5) has the higher relative contact rate.}
    \label{fig:transmission_var}
\end{figure}

Figure \ref{fig:transmission_var} shows how the final attack rate of a epidemic varies with the probability of transmission, $q$. When varying transmission probability, we vary the transmission probability of all ethnicity groups at the same time. $R_0$ varies linearly with transmission probability, the slope of the scenarios is the value of the largest eigenvalue when $q=1$. The threshold for a epidemic ($R_0>1$) to occur is therefore different for each scenario. Variation in population wide dynamics between scenarios is not very pronounced; the variation between ethnicity groups in difference scenarios is significantly greater. As a consequence of differences in the size and relative contact rates of different ethnicity groups, scenarios 1 and 2 have a lower attack rate for both ethnicity groups (compared to the other scenarios) while having a similar population-wide attack rate. 

\subsubsection{Ethnic assortativity variation}
\begin{figure}[ht]
    \centering
    \includegraphics[width=0.99\linewidth]{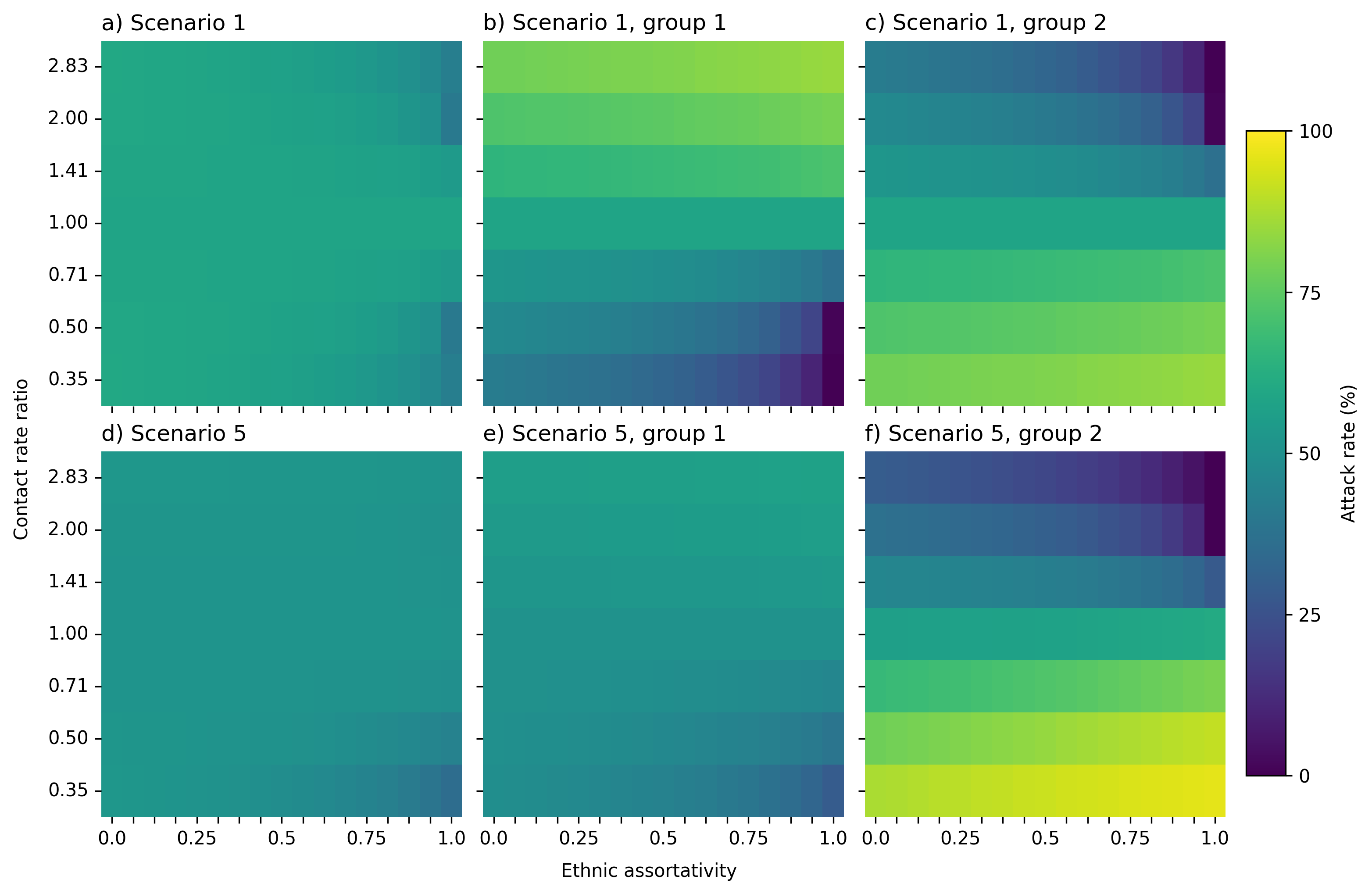}
    \caption{Heat-maps showing how variation in the ethnic assortativity and contact rate ratio affects the attack rate for a) the whole population of scenario 1; b) the population of ethnicity group 1 in scenario 1; c) the population of ethnicity group 2 in scenario 1; d) the whole population of scenario 5; e) the population of the majority ethnicity group in scenario 5; f) the population of the minority ethnicity group in scenario 5.}
    \label{fig:heatmap_eth_ep_and_rel_contact_rate}
\end{figure}

Figure \ref{fig:heatmap_eth_ep_and_rel_contact_rate} shows the final epidemic size in scenario 1 and 5 when varying the ethnicity group's assortativity and relative contact ratio. We focus on these scenarios as the results from other scenarios do not significantly differ from one the presented scenarios (Scenario 2 results are similar to scenario 1, while results from scenarios 3 and 4 are similar to scenario 5). From these plots it can be seen that assortativity can have a relatively small effect on overall final epidemic size when one group is significantly larger than the other, or zero impact on final epidemic size when relative contact rates are equal (and each group has the same initial fraction infected). This shows that, if the model were fitted to attack rate data, a comparable quality of fit could be obtained by different parameter combinations. This implies that  these parameters are likely to be non-identifiable from attack rate data.

\subsection{POLYMOD projection results}

\begin{figure}[ht]
    \centering
    \includegraphics[width=0.7\linewidth]{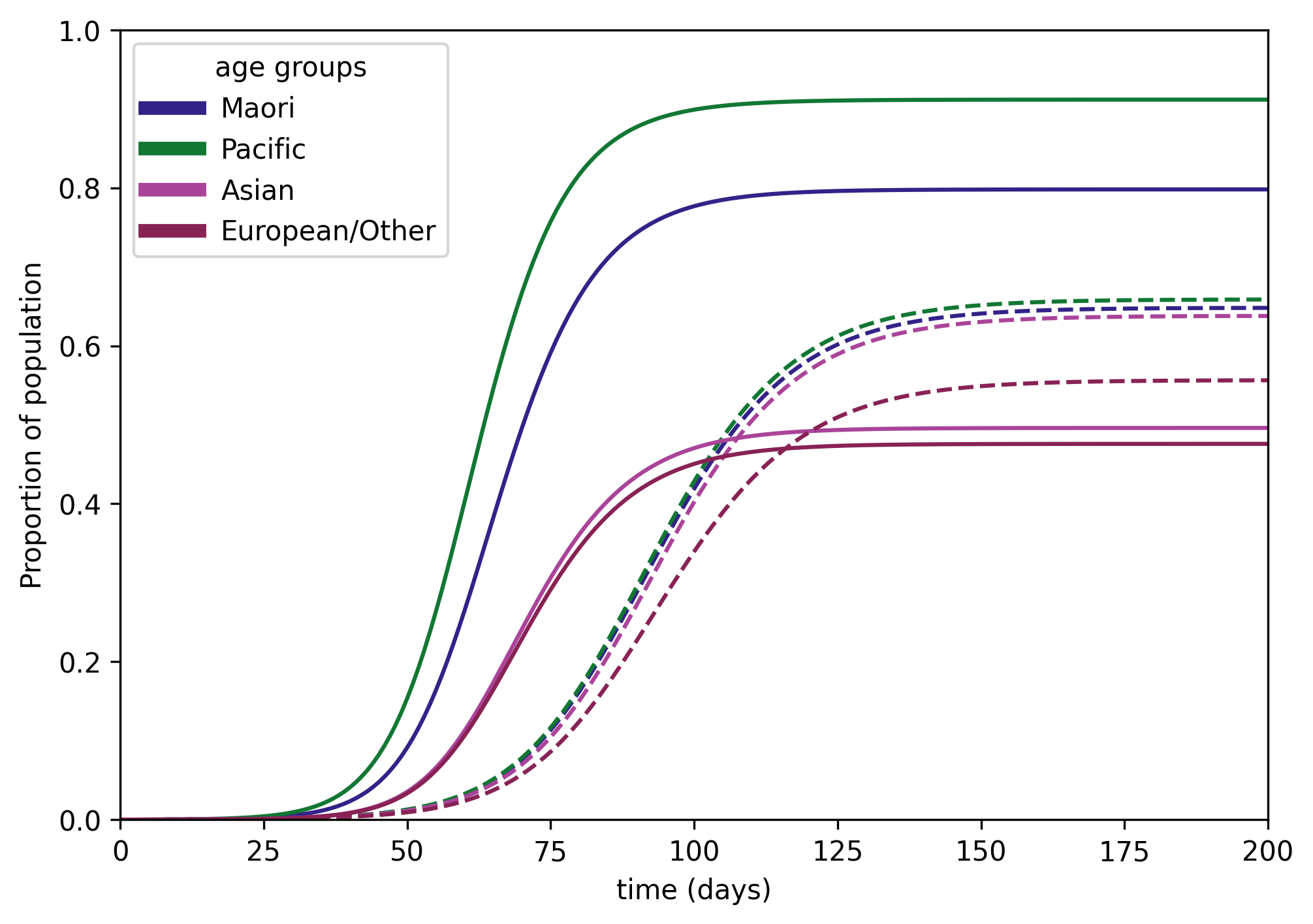}
    \caption{Ethnicity-aggregated results of a SEIR model run using the POLYMOD contact survey projected onto Aotearoa New Zealand. The solid lines represent the results for the ethnicity groups having different relative contact rates (those being $F_a=2$, $3$, $0.9$, and $1$ for Māori, Pacific, Asian, and European/Other people, respectively) and the dashed lines represent the results for all ethnicity groups having the same relative contact rates ($F_a=1$ for all ethnicity groups). Ethnic assortativity is set to $0.2$ for this simulation, the disease development rate is set to be $\sigma=1\text{ day}^{-1}$, and the recovery rate is set as $\gamma = 2/3\text{ day}^{-1}$.}
    \label{fig:eth_POLYMOD}
\end{figure}

Figure \ref{fig:eth_POLYMOD} shows that despite the contact rate ratio of Asian to European people being 0.9, European people saw fewer cases in our simulation. The younger age structure of Asian individuals compared to European individuals, with Asian people having a median age of 33.8 compared to the European median age of 41.7 \cite{statsnz_place_2023}, caused the groups to have similar average population-wide contact rates. The addition of relative contact rates also affected both the final epidemic state and shortened the amount of time it takes for the epidemic to end (This can be seen in Appendix D).

The extension of the social contact matrix with equal relative contact rates is equivalent to running a model without considering ethnicity and then doing a weighted average using the age distribution within the ethnicity groups. This is an implicit assumption other models make when when discussing potential ethnicity-based differences from an age-based model.

An age-only model may slightly overestimate the overall number of cases (Appendix D). When we accounted for ethnicity, there was a roughly 2\% decrease in the population of Aotearoa New Zealand infected, which is roughly 100,000 people. The effectiveness of different policy decisions may be affected when you look at total population effects. Adding ethnicity increases the attack rate of the older population while decreasing it for the younger population, which would be important for diseases where severity was significantly worse in a specific age group such as youth or elderly. In the younger generation, a high proportion of the group was already getting infected in the age-only model and so the increase in social contact rates for Māori and Pacific populations (which are relatively young) had a smaller impact than the reduction in contact rates for Asian and European/Other populations (which are older). In the older groups the opposite was the case, which caused slight increases in the older population's attack rates.

\section{Discussion}

We presented a method to extend a given social contact matrix with additional socio-demographic factors, specifying age and ethnicity as an example of two factors of interest. We did this in a way that did not require estimates of pairwise social interaction rate between each age-ethnicity group and instead only needed information on the the extent to which mixing is assortative with respect to ethnicity and the relative contact rate of each ethnicity group.

We then analysed a hypothetical epidemic under some simple population structures and then over a population representing Aotearoa New Zealand to analyse obscured variation in attack rates, reproduction numbers, and inequities. We found that differences in relative contact rates between ethnicity groups and assortative mixing patterns can lead to substantial disparities in attack rates. The magnitude of these disparities depends on the relative sizes of the groups, as well as their age structure and age-based mixing patterns. 

\subsection{Strengths and limitations}

The main strength of our method is that is allows infectious disease models to stratify outputs by additional socio-demographic factors, including in situations where fine-grained data on contact rates between each combination of socio-demographic groups are lacking. This is important because it enables models to quantify the effects of systematic differences between groups in factors such as age structure, vaccination rates and contact rates on disease impact.  

The main limitation our method has is the imposed structure on the social contact matrix. Our method allows the construction of an extended social contact matrix with limited information being required, however the lack of information means we need to make assumptions about the structure of group interactions (like assortative mixing). As a simplifying assumption, we assumed that the degree of assortativity was the same for all ethnicity groups and all age groups. However, in some contexts there is evidence for greater heterogeneity within and between age groups as people get older \cite{kuchel_workshop_2025}. This could imply that older individuals would have relatively higher assortativity values, which our model cannot account for. A similar effect may be present in other socio-demographic factors. Within Aotearoa New Zealand, Asian and Pacific People have high proportions of their populations living in the Auckland region while Māori and European/Other people are more dispersed around the country \cite{statsnz_place_2023}. We would expect that Asian and Pacific people would therefore have higher levels of assortativity compared to other ethnicity groups as there is higher clustering in those ethnicity groups. An agent-based model has shown how racial inequities in the USA arise from household size and workplace settings, which would also depend on spatial clustering \cite{harris_why_2025}.

The contact rate ratios defined the difference in the average contact rate for two comparable groups. For example, two ethnicity groups in the same age group would have an average difference in contact rate defined by the ratio of their group's relative contact rate. In other terms, the contact rate ratio is not the average population-wide contact ratio between two ethnicity groups (as this would also depend on the age structure). Similar to assortativity, we assume that relative contact rates are independent of age. This is a reasonable assumption in the absence of fine-grain age-specific data, but this may not always be accurate.

Depending on the choice of $G_{ija}$, our method can result in negative social contact matrix elements when the relative contact rates and population structures of the socio-demographic factors lead to extreme differences in the distribution of contacts and there is a high assortativity value. Before matrices resulting from our method are used, non-negativity of matrix elements should be verified. Our method provides a way to define an extended contact matrix that satisfies necessary consistency conditions while allowing ethnicity-specific contact rates to be specified and may allow exploration of different epidemic dynamic depending on the choice of $G_{ija}$. However, other ways of formulating the matrix may exist which could also lead to different epidemic dynamics. Possible avenues for future work could be to explore the impact of alternative ways of extending the contact matrix on epidemic dynamics and to generalise our derivation to allow for heterogeneity in the additional socio-demographic factor to depend on one, or both, of the socio-demographic factors instead of being independent of both.

Our model also has the same general limitations inherent to compartment-based models and social contact matrices. One issue with the use of social contact surveys can be perception bias issues which can lead to inaccurate social contact rates \cite{harris_simulating_2025}. The probability of infection resulting from each contact is difficult to know and, in the absence of data, we assume that this is the same for all contacts, which may not be the case. Ideally, contacts would be weighted by their probability of infection, but it is not feasible to find the probability of infection resulting from each interaction. Compartmental models assume that the population is homogeneous and well-mixed within each compartment and often do not take spatial effects into account. We expect the spread of an infectious disease to be be greater in urban environments and covariance in social contacts to reduce the total number of infected people.

\subsection{Effect of ethnic heterogeneity on model outputs}

The extended model showed that there can be differences in group-specific disease incidence even if the aggregate remains the same. The aggregate disease incidence can also change as a result of ethnic heterogeneity, but the aggregate information alone cannot uniquely determine the inequity present. The variation in attack rate and basic reproductive number that are obscured by ignoring variation in ethnicity, but more generally socio-demographic factors, can affect epidemic health outcomes and preventative policy decisions. An increase in the attack rate would lead to a higher rate of severe disease and mortality than other models would expect, while an increase in the basic reproductive number would increase the rate of epidemic growth at the early stages of the epidemic. This could lead to hospitals not having enough patient capacity even with a reduction in the overall case count.

The effect of variation in both relative contact rate and assortativity is more pronounced in the smaller population groups. This means that including these parameters in models can significantly influence the attack rates seen in minority groups, which are often the most affected in epidemic scenarios. An equivalent statement is that variation in minority group health outcome can be obscured by using population averages.

In our analysis of a projected age-based contact matrix, the addition of ethnicity as a factor had three main effects on model outputs: the basic reproductive number increased; the total number of infections decreased; and the age structure of infections shifted from younger to older individuals. These three effects can all be critical to effectively inform policy decisions, particularly during an epidemic.

\section{Conclusion}

Our method allows the extension of social contact matrices with additional socio-demographic factors. Our approach is applicable in contexts where detailed socio-demographic information from contact surveys is not available and can serve as an inexpensive estimate. The extended social contact matrix enables the development of models that quantify epidemic dynamics and outcomes in groups of interest. Such models are better equipped to inform policy as they enable greater consideration of socio-demographic inequities in infectious disease impact, as well as the effect of interventions designed to minimise inequalities. 

\section{Code and data availability}
All code and data used for this project can be found at \url{https://github.com/Vincentlomas/Age_ethnicity_stratification} with an archive available \cite{lomas_vincentlomasage_ethnicity_stratification_2026}.

\section*{Disclosure statement}

We declare that we have no potential conflicts of interest with respect to the investigation, authorship, and publication of this article.

\section*{Funding}
The project ``Improving models for epidemic preparedness and response: modelling differences in infectious disease dynamics and impact by ethnicity'' (TN/P/24/UoC/MP) was funded by Te Niwha, the Infectious Diseases Research Platform – co-hosted by PHF Science and the University of Otago and provisioned by the Ministry of Business, Innovation and Employment, New Zealand. This research was supported by the Marsden Fund grant (24-UOC-020) managed by Royal Society Te Apārangi.

\printbibliography

\section*{Appendix A: Satisfaction of conditions}
Here we present proof of satisfaction of conditions for the presented matrices. We do not use the $\bar F$ notation used in the main body of text and instead present the matrices in terms of $N$, $F$, and $C_{ij}$.

\subsection*{Proportionate mixing}
To gain intuition into the definition of the proportionate mixing matrix in Eq. (5) of the main text, observe that the total number of contacts between individuals in age group $i$ and individuals in age group $j$ is $N_iC_{ij}$. Similarly, the total number of contacts between individuals in age group $i$ and ethnicity group $a$ and individuals in age group $j$ and ethnicity group $b$ is $N_{ia}C_{ia,jb}$. Hence, if we distribute the total contacts between age groups $i$ and $j$ across ethnicity groups in proportion to the fraction of age groups $i$ and $j$ in those groups ($N_{ia}/N_i$ and $N_{jb}/N_j$ respectively) and their relative contact rates ($F_a$ and $F_b$), we obtain
\[
N_{ia}C_{ia,jb}^{\text{proportionate}} = k N_i C_{ij} \frac{N_{ia}N_{jb}}{N_i N_j} F_a F_b
\]
where $k$ is an age-specific normalisation constant. As will be shown below, the aggregation constraint requires that
\[
k = \frac{N_i N_j}{  \sum_{a'}F_{a'}N_{ia'}\sum_{b'}F_{b'}N_{jb'} }
\]
Dividing the first equation above by $N_{ia}$ results in the expression for $C_{ia,jb}^{\text{proportionate}}$ in Eq. (5) of the main text.

To check the symmetry constraint, see that

\begin{align*}
    N_{ia}C^{\text{proportionate}}_{ia,jb} = \frac{F_aF_bC_{ij}N_iN_{ia}N_{jb}}{\sum_{a'}F_{a'}N_{ia'}\sum_{b'}F_{b'}N_{jb'}}\\
    N_{jb}C^{\text{proportionate}}_{jb,ia}  = \frac{F_aF_bC_{ji}N_jN_{ia}N_{jb}}{\sum_{a'}F_{a'}N_{ia'}\sum_{b'}F_{b'}N_{jb'}}
\end{align*}

These are equal as a consequence of the symmetry condition on the primary contact matrix ($C_{ij}N_i=C_{ji}N_j$).

To check the aggregation constraint, see that

\begin{align*}
    \frac{\sum_{a,b}N_{ia}C^{\text{proportionate}}_{ia,jb}}{N_{i}} & = \frac{C_{ij}N_i\sum_{a,b}F_aF_bN_{ia}N_{jb}}{N_i\sum_{a'}F_{a'}N_{ia'}\sum_{b'}F_{b'}N_{jb'}}\\
    & = C_{ij}
\end{align*}

To check the relative interaction constraint, see that

\begin{align*}
    \frac{\sum_{j,b}C^{\text{proportionate}}_{ia_1,jb}}{\sum_{j,b}C^{\text{proportionate}}_{ia_2,jb}} &=\frac{\sum_{j,b}\left(\frac{F_{a_1}F_bC_{ij}N_iN_{jb}}{\sum_{a'}F_{a'}N_{ia'}\sum_{b'}F_{b'}N_{jb'}}\right)}{\sum_{j,b}\left(\frac{F_{a_2}F_bC_{ij}N_iN_{jb}}{\sum_{a'}F_{a'}N_{ia'}\sum_{b'}F_{b'}N_{jb'}}\right)}\\
    & = \frac{F_{a_1}\sum_{j,b}\left(\frac{F_bC_{ij}N_iN_{jb}}{\sum_{a'}F_{a'}N_{ia'}\sum_{b'}F_{b'}N_{jb'}}\right)}{F_{a_2}\sum_{j,b}\left(\frac{F_bC_{ij}N_iN_{jb}}{\sum_{a'}F_{a'}N_{ia'}\sum_{b'}F_{b'}N_{jb'}}\right)}\\
    & = \frac{F_{a_1}}{F_{a_2}}
\end{align*}

Note that this matrix recovers the equal contact frequency proportionate mixing matrix when the frequencies for the additional socio-demographic groups are set to be equal.

\subsection*{Segregated mixing}
The segregated mixing matrix is defined by the following:
\begin{align*}
C^{\text{segregated}}_{ia,jb}&= \begin{cases}
        \delta_{ab}\left[\frac{F_a}{\bar F_i} \sum_{k} C_{ik} - \frac{N_i}{N_{ia}} \sum_{k\neq i} C_{ik} G_{ika}\right] &\text{if } i=j,\\
        \delta_{ab}\frac{N_iC_{ij}}{N_{ia}}  G_{ija} & \text{otherwise},
    \end{cases}
\end{align*}
where $G_{ija}=G_{jia}$ and $\sum_a G_{ija} = 1$.
Or equivalently,
\begin{align*}
    C^{\text{segregated}}_{ia,jb}&= \begin{cases}
        \delta_{ab}\left[\frac{F_aC_{ij}}{\bar F_i} +\sum_{k\neq i}\left(\frac{F_aC_{ik}}{\bar F_i}-\frac{N_iC_{ik}}{N_{ia}}  G_{ika}\right)\right] &\text{if } i=j,\\
        \delta_{ab}\frac{N_iC_{ij}}{N_{ia}}  G_{ija} & \text{otherwise},
    \end{cases} \label{eq:seg_matrix}
\end{align*}
\newline
To check the symmetry constraint, first note that the within age-group interaction section of the matrix ($i=j$) is necessarily symmetric due to only having non-zero values when $a=b$. When analysing $i\neq j$, the following holds
\begin{align*}
    N_{ia}C^{\text{segregated}}_{ia,jb} &= \delta_{ab}N_iC_{ij}  G_{ija}\\
    N_{jb}C^{\text{segregated}}_{jb,ia}  &= \delta_{ba}N_jC_{ji}  G_{jia}
\end{align*}
These are equal as a consequence of the symmetry condition on the age-stratified contact matrix ($C_{ij}N_i=C_{ji}N_j$) and the general function ($G_{ija}=G_{jia})$ and that the matrix has non-zero values only when $a=b$.
\newline
To check the aggregation constraint, see that when $i=j$
\begin{align*} 
    \frac{\sum_{a,b}N_{ia}C^{\text{segregated}}_{ia,jb}}{N_{i}} & =\frac 1{N_i}\sum_{a,b}\delta_{ab}N_{ia}\left[\frac{F_aC_{ij}}{\bar F_i} +\sum_{k\neq i}\left(\frac{F_aC_{ik}}{\bar F_i}-\frac{N_iC_{ik}}{N_{ia}}  G_{ika}\right)\right]\\
    & = \frac 1{N_i}\sum_{a}N_{ia}\left[\frac{F_aN_iC_{ij}}{\sum_{a'}F_{a'}N_{ia'}} +\sum_{k\neq i}\left(\frac{F_aN_iC_{ik}}{\sum_{a'}F_{a'}N_{ia'}}-\frac{N_iC_{ik}}{N_{ia}}  G_{ika}\right)\right]\\
    & = \frac 1{N_i}\left[\frac{N_iC_{ij}\sum_{a}F_aN_{ia}}{\sum_{a'}F_{a'}N_{ia'}} +\sum_{k\neq i}\left(\frac{N_iC_{ik}\sum_{a}F_aN_{ia}}{\sum_{a'}F_{a'}N_{ia'}}-N_iC_{ik}\sum_{a}G_{ika}\right)\right]\\
    & = \frac 1{N_i}\left[N_iC_{ij} +\sum_{k\neq i}\left(N_iC_{ik}-N_iC_{ik}\right)\right]\\
    & = C_{ij}
\end{align*}
When $i\neq j$
\begin{align*} 
    \frac{\sum_{a,b}N_{ia}C^{\text{segregated}}_{ia,jb}}{N_i} & = \frac 1 {N_i}\sum_{a,b}\delta_{ab}N_{ia}\frac{N_iC_{ij}}{N_{ia}} G_{ija}\\
    & = C_{ij} \sum_a G_{ija}\\
    & = C_{ij}
\end{align*}
To check the relative interaction constraint, see that
\begin{align*}
    \sum_{j,b}C^{\text{segregated}}_{ia,jb}
    & =\sum_{b}\left(\delta_{ab}\left[\frac{F_aC_{ii}}{\bar F_i} +\sum_{k\neq i}\left(\frac{F_aC_{ik}}{\bar F_i}-\frac{N_iC_{ik}}{N_{ia}}  G_{ika}\right)\right] +\sum_{j\neq i}\delta_{ab}\frac{N_iC_{ij}}{N_{ia}}  G_{ija}\right)\\
    & =\frac{F_aC_{ii}}{\bar F_i} +\sum_{k\neq i}\frac{F_aC_{ik}}{\bar F_i}-\sum_{k\neq i}\frac{N_iC_{ik}}{N_{ia}}  G_{ika} +\sum_{j\neq i}\frac{N_iC_{ij}}{N_{ia}}  G_{ija}\\
    & = F_a\sum_{k}\frac{C_{ik}}{\bar F_i}
\end{align*}
Therefore
\begin{align*}
    \frac{\sum_{j,b}C^{\text{segregated}}_{ia_1,jb}}{\sum_{j,b}C^{\text{segregated}}_{ia_2,jb}} & = \frac{ F_{a_1}\sum_{k}\frac{C_{ik}}{\bar F_i}}{F_{a_2}\sum_{k}\frac{C_{ik}}{\bar F_i}}\\
    & = \frac{F_{a_1}}{F_{a_2}}
\end{align*}
In this study, we assumed $G_{ija}$ takes the form
\begin{align*}
    G_{ija}=\frac12\left(\frac{F_aN_{ia}}{N_i\bar F_i}+\frac{F_aN_{ja}}{N_j\bar F_j}\right)
\end{align*}
Some other possible choices we considered for $G_{ija}$ were the following:
\begin{align*}
    G_{ija} & = \frac{F_a(N_{ia}+N_{ja})}{N_{i}\bar{F}_i + N_j\bar{F}_j }\\
    \text{or}\quad G_{ija} &=  \frac{F_a\sqrt{N_{ia}N_{ja}}}{\sum_{a'} F_{a'}\sqrt{N_{ia'}N_{ja'}}}
\end{align*}

\section*{Appendix B: Population structures}
In Table \ref{tab:scenario_table_full} present the hypothetical population structures used for numerical simulation. 

\begin{table}[ht]
    \centering
    \begin{tabular}{c|c|c}
        Scenario & Population structure (1000s) & Age group contact rates\\
        \hline
        1 &$\begin{bmatrix}
            30 & 30 & 30 & 30 & 30\\
            30 & 30 & 30 & 30 & 30
        \end{bmatrix}^T$& $\begin{bmatrix}
            1&1&1&1&1
        \end{bmatrix}$ \\
        \hline
        2 & $\begin{bmatrix}
            20 & 25 &30 & 35 & 40\\
            40 & 35 & 30 & 25 & 20
        \end{bmatrix}^T$ & $\begin{bmatrix}
            1&1&1&1&1
        \end{bmatrix}$\\
        \hline
        3 & $\begin{bmatrix}
            54 & 54 &54 &54 &54  \\
            6& 6& 6& 6& 6
        \end{bmatrix}^T$ & $\begin{bmatrix}
            1&1&1&1&1
        \end{bmatrix}$\\
        \hline
        4 & $\begin{bmatrix}
            36 & 45 & 54 & 63 & 72\\
            8 & 7 & 6 & 5 & 4
        \end{bmatrix}^T$ & $\begin{bmatrix}
            1&1&1&1&1
        \end{bmatrix}$\\
        \hline
        5 & $\begin{bmatrix}
            36 & 45 & 54 & 63 & 72\\
            8 & 7 & 6 & 5 & 4
        \end{bmatrix}^T$ & $\frac{15}{86}\begin{bmatrix}
            8 & 7 & 6 & 5 &4
        \end{bmatrix}$
    \end{tabular}
    \caption{Table of population structures and contact rates for the five scenarios considered. The population structure has five age groups and two ethnicity groups; the $i^\text{th}$ row and $a^\text{th}$ column of the population matrices show the population size in age group $i$ and ethnicity group $a$. Scenario 5's age rates were chosen to be linear with age group with the oldest age group having have the contact rate compared to the youngest group; the rates were then normalised such that the total number of contacts in the population remained the same.}
    \label{tab:scenario_table_full}
\end{table}

\section*{Appendix C: Results when holding the basic reproduction number constant}
\begin{figure}[ht]
    \centering
    \includegraphics[width=0.95\linewidth]{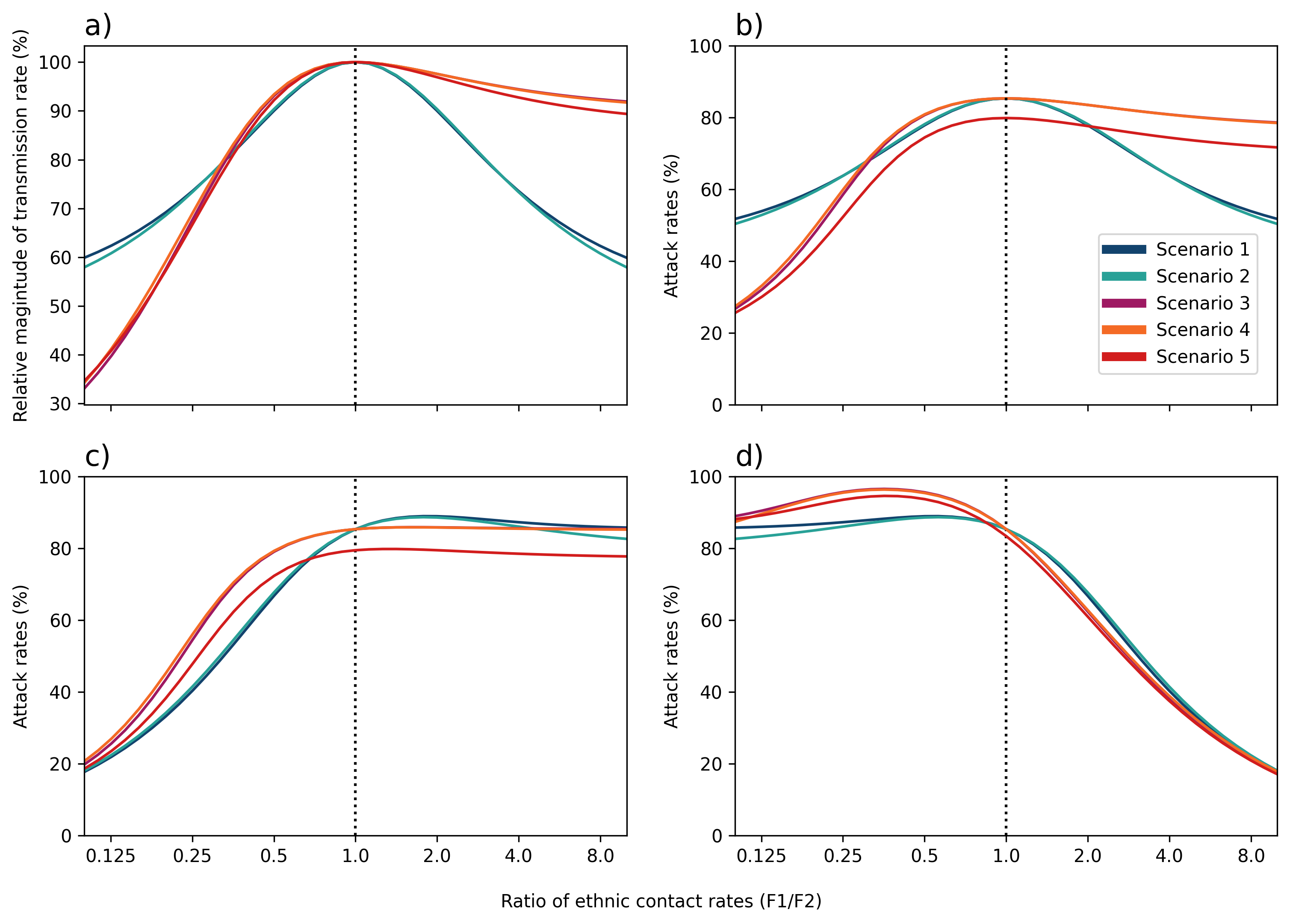}
    \caption{Plots of the variation in a) the number of social contacts retained; b) the whole population attack rates; c) ethnicity group 1's attack rates; and d) ethnicity group 2's attack rates with relative contact rate. A lower contact rate ratio means the ethnicity group with the lower population size (if present) has the higher contact rate. The ethnic assortativity is 0.3.}
    \label{fig:const_R0_age_POLYMOD}
\end{figure}

\begin{figure}[ht]
    \centering
    \includegraphics[width=0.99\linewidth]{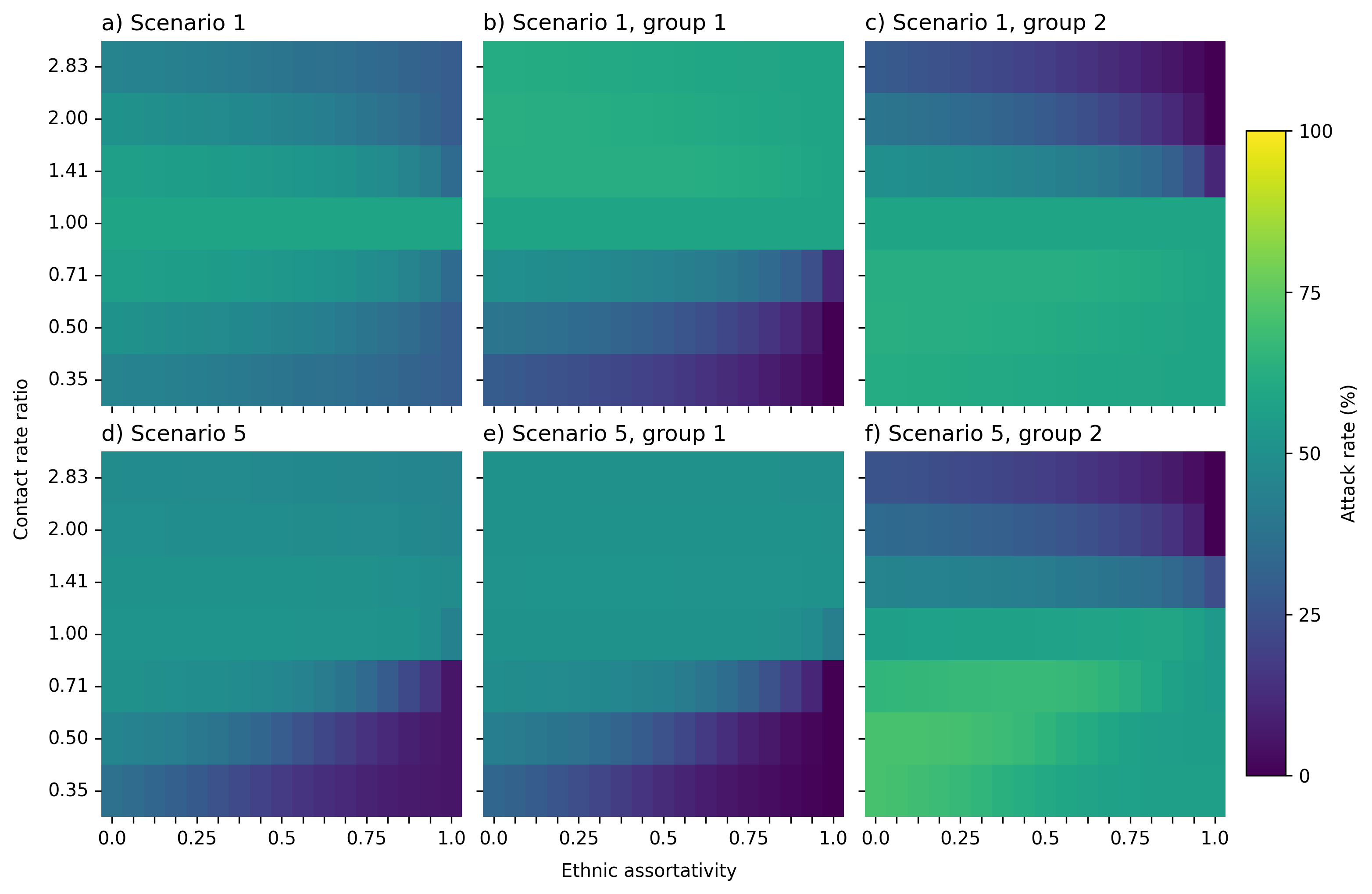}
    \caption{Heat-maps showing how variation in the ethnic assortativity and contact rate ratios affects the attack rate for a) the whole population of scenario 1; b) the population of ethnicity group 1 in scenario 1; c) the population of ethnicity group 2 in scenario 1; d) the whole population of scenario 5; e) the population of the majority ethnicity group in scenario 5; f) the population of the minority ethnicity group in scenario 5.}
    \label{fig:const_R0_heatmap_eth_ep_and_rel_contact_rate}
\end{figure}

Here we present results for contact matrix extensions where the basic reproduction number was held constant. The basic reproduction number was held constant by extending the matrix as described in the main text and then multiplying the matrix by a constant (shown in Figure \ref{fig:const_R0_age_POLYMOD}a) such that the basic reproduction number was $1.5$. Figures \ref{fig:const_R0_age_POLYMOD} and \ref{fig:const_R0_heatmap_eth_ep_and_rel_contact_rate} show how variation in contact rate ratio and ethnic assortativity affects final attack rates of an outbreak. The overall population attack rate decreases as the contact rate ratio deviates from 1. The minority group no longer sees the same magnitude of infections that they saw when the basic reproductive number was not held constant. Ethnicity groups no longer approach attack rates of 100\% and instead see a reduction of attack rates when assortativity approaches 1.
\newline

\section*{Appendix D: Age aggregation of models}
\begin{figure}[ht]
    \centering
    \includegraphics[width=0.95\linewidth]{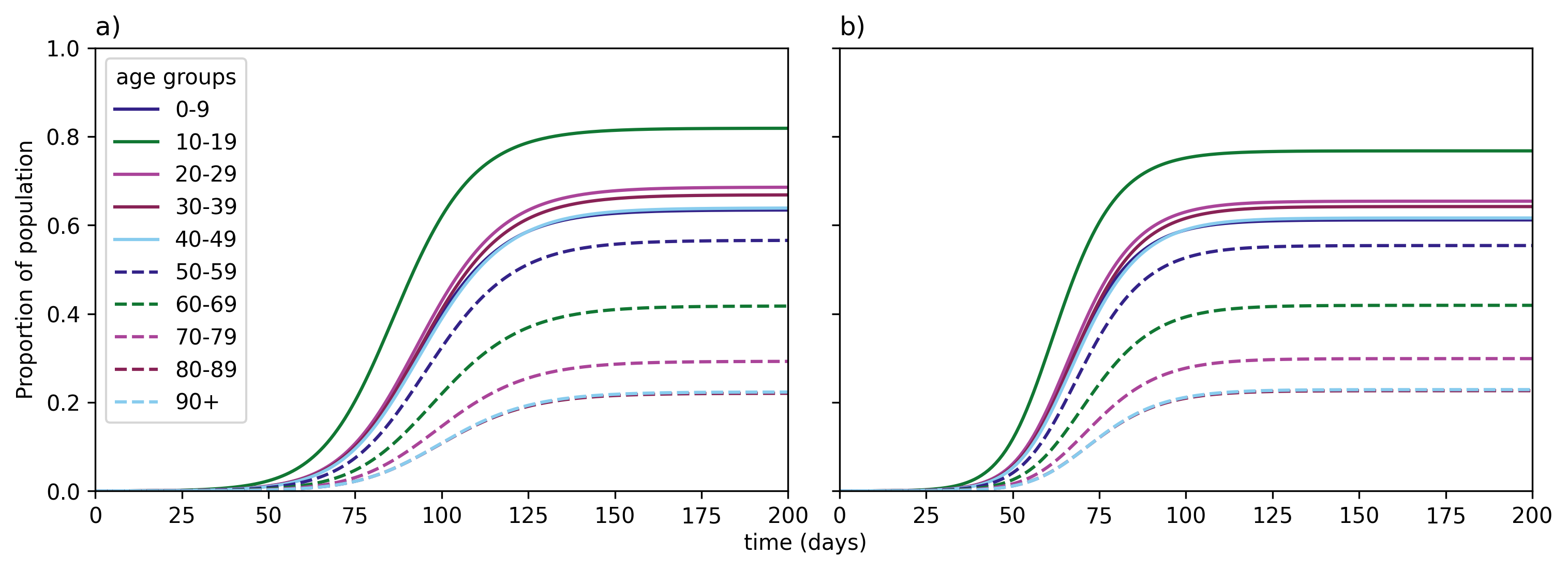}
    \caption{Age-aggregated results of a SEIR model run using the POLYMOD contact survey projected onto Aotearoa New Zealand with a) no information about ethnicity and b) ethnicity added as a factor into the social contact matrix with an assortativity value of $20\%$ and relative contact rates of 2, 3, 0.9, and 1 for Māori, Pacific, Asian, and European/Other groups, respectively.}
    \label{fig:age_POLYMOD}
\end{figure}

\begin{figure}[ht]
    \centering
    \includegraphics[width=0.7\linewidth]{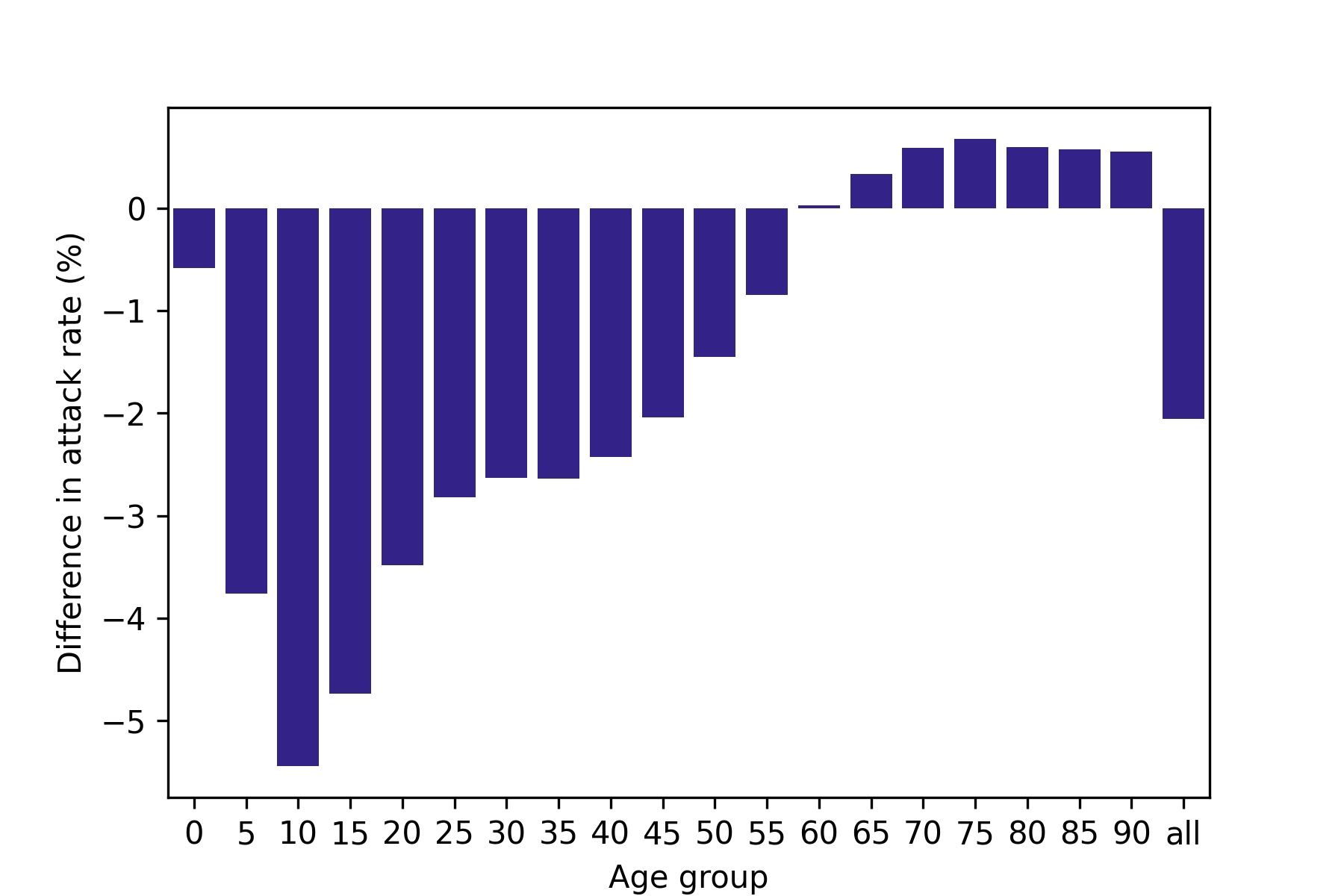}
    \caption{Bar-plot showing the percentage difference (of the population) in final attack rates when ethnicity is added to the model. Age contact matrix was determined by projecting the POLYMOD contact matrix onto Aotearoa New Zealand. Extended contact matrix had relative contact rates of $F_a=2$, $3$, $0.9$ and $1$ for Māori, Pacific, Asian, and European/Other groups, respectively, and an ethnic assortativity of 0.2.}
    \label{fig:barplot_POLYMOD}
\end{figure}

In figure \ref{fig:age_POLYMOD} we show how the addition of ethnicity affects the age-based results of a simulated epidemic. The epidemic occurs sooner due to the inclusion of heterogeneity increasing the basic reproductive number. The final attack rates are relatively similar with slight age-dependence caused by population structures shown in Figure \ref{fig:barplot_POLYMOD}. The magnitude of these differences will increase or decrease not only with the choices of parameter values when using our method, but also with the final attacks rates before the extension of our model. For example a disease with relatively low infectivity will see small changes as will a disease with relatively high infectivity.

\end{document}